# Mitigating MRI Domain Shift in Sex Classification: A Deep Learning Approach with ComBat Harmonization

1st Peyman Sharifian
Department of Biomedical Engineering
University of Isfahan
Isfahan, Iran
peyman.sharifian.1996@gmail.com

2nd Mohammad Saber Azimi
Department of Medical Radiation Engineering,
Shahid Beheshti University,
Tehran, Iran,
saber.azimi2020@gmail.com

3rd Prof. Ali Reza Karimian
Department of Biomedical Engineering, University of Isfahan
Isfahan, Iran
karimian@eng.ui.ac.ir

4th Dr. Hossein Arabi
Division of Nuclear Medicine & Molecular Imaging
Geneva University Hospital,
Geneva, Switzerland
hossein.arabi@unige.ch

**Abstract—** Deep learning models for medical image analysis often suffer from performance degradation when applied to data from different scanners or protocols, a phenomenon known as domain shift. This study investigates this challenge in the context of sex classification from 3D T1-weighted brain magnetic resonance imaging (MRI) scans using the IXI and OASIS3 datasets. While models achieved high within-domain accuracy (around 0.95) when trained and tested on a single dataset (IXI or OASIS3), we demonstrate a significant performance drop to chance level (about 0.50) when models trained on one dataset are tested on the other, highlighting the presence of a strong domain shift. To address this, we employed the ComBat harmonization technique to align the feature distributions of the two datasets. We evaluated three state-of-the-art 3D deep learning architectures (3D ResNet18, 3D DenseNet, and 3D EfficientNet) across multiple training strategies. Our results show that ComBat harmonization effectively reduces the domain shift, leading to a substantial improvement in cross-domain classification performance. For instance, the cross-domain balanced accuracy of our best model (ResNet18 3D with Attention) improved from approximately 0.50 (chance level) to 0.61 after harmonization. t-SNE visualization of extracted features provides clear qualitative evidence of the reduced domain discrepancy post-harmonization. This work underscores the critical importance of domain adaptation techniques for building robust and generalizable neuroimaging AI models.

## I. Introduction

The application of deep learning to neuroimaging has shown remarkable success in tasks such as disease diagnosis, prognosis, and biomarker discovery [1-3]. However, a significant obstacle to the clinical deployment of these models is their susceptibility to domain shift [4]. Domain shift occurs when a model trained on data from a specific source (for example, a particular MRI scanner, acquisition protocol, or patient population) experiences a drop in performance when applied to data from a different target domain. This variability introduces confounding effects that can overshadow the actual biological signals of interest, a problem starkly evident when models trained on public datasets like IXI fail to generalize to clinical cohorts like OASIS3. This performance degradation highlights that without explicit strategies to learn scanner-invariant features [5], models will inevitably exploit these spurious correlations, compromising their clinical utility. To address this challenge and bridge the domain gap, data harmonization techniques have emerged as a crucial preprocessing step to enable the development of robust and generalizable models.

Sex classification from brain MRI serves as a robust benchmark task for evaluating neuroimaging models because of the well-established neuroanatomical differences between males and females [6]. While models can achieve near-perfect accuracy within the same dataset (same domain), their performance often collapses on external datasets (cross-domain). This performance degradation is not merely an anecdotal observation but a systematic vulnerability; recent benchmarking platforms have quantitatively demonstrated that modern deep neural networks are highly susceptible to specific distribution shifts caused by variances in scanners and acquisition protocols, which are endemic to multi-site neuroimaging research [7]. This reveals the model's reliance on scanner-specific artifacts rather than generalizable anatomical features. Therefore, the sex classification task provides a controlled and measurable paradigm to quantify the extent of domain shift and rigorously evaluate the efficacy of proposed harmonization techniques against a known biological ground truth.

Deep learning is now a widely used tool in recent studies aiming to determine sex from brain scans, showcasing its significant potential alongside its notable limitations. Research using structural MRI has proven highly effective, with one example being a 3D convolutional neural network that reached 87% accuracy on four distinct datasets with little preprocessing. However, this study also underscored a major vulnerability: without meticulous dataset design, models can easily learn to rely on simple, confounding factors like overall brain size rather than more meaningful patterns [6]. This specific issue was tackled in another investigation where subjects were carefully matched for overall head size

(intracranial volume). Despite this control, the model still surpassed 92% accuracy. This strongly suggests that neuroanatomical sex differences are not merely about size but are complex, involving multiple subtle and nuanced features [8]. To capture more refined biological markers, scientists have turned to other imaging techniques. For example, a study employing diffusion MRI (which visualizes brain wiring) with advanced AI models achieved outstanding results (AUC 0.92-0.98), pinpointing major white matter pathways as crucial for differentiation and implying underlying microstructural variations[9]. Similarly, research showed that DTI-based mean diffusivity in the substantia nigra can discriminate Parkinson's patients from healthy controls, underscoring the broader role of diffusion metrics and machine learning in neuroimaging [10]. Separately, research on fMRI (which measures brain activity) created an innovative method using stochastic encoding and an ensemble of CNNs to classify sex (AUROC around 0.85), highlighting the importance of specific functional networks related to attention and executive function, all while using a technique to minimize the influence of confounding factors [11]. The element of brain development over time has also been examined. A model that analyzed scans of adolescents from the ABCD study over time achieved remarkable accuracy (over 97%) and discovered that its ability to distinguish sex improved with the subjects' age, connecting these anatomical shifts to the biological processes of puberty [12]. The utility of deep learning for sex classification isn't limited to the brain. Its adaptability is demonstrated by its successful application to a completely different anatomical feature—the hyoid bone in the neck. Using a specialized architecture for 3D data, a model achieved 88.71% accuracy, showcasing its value for forensic identification [13]. Despite these successes, a critical and pervasive weakness emerges in almost every study: a severe susceptibility to domain shift. This problem was made abundantly clear by research showing that a model's performance catastrophically declined when it was applied to data acquired from a different brand of MRI scanner. Most importantly, this same research revealed that common data standardization methods, such as ComBat, were insufficient to fix this problem. This finding exposes a fundamental barrier that prevents these highly accurate models from being reliably used in actual clinical settings [14].

In this study, we explicitly quantify this domain shift using the IXI and OASIS3 datasets. We then address it using ComBat, a popular harmonization method initially developed for genomic data and later adapted for neuroimaging [15]. ComBat adjusts for batch effects by standardizing the mean and variance of features across different datasets (batches) while preserving biological variances of interest. This work makes three key contributions. First, we demonstrate that while modern 3D deep learning models achieve expert-level accuracy on within-domain sex classification, they fail catastrophically under cross-domain conditions, quantitatively confirming a severe domain shift between the IXI and OASIS3 datasets. Second, we show that ComBat harmonization effectively mitigates this issue, significantly improving cross-domain performance. Finally, we provide a multi-faceted validation: quantitative metrics show a major boost in balanced accuracy, while t-SNE visualizations offer qualitative proof of successful feature distribution alignment post-harmonization.

## II. MATERIALS AND METHODS

### A. Datasets

Our study utilized two distinct publicly available neuroimaging datasets, IXI and OASIS3 [16], to investigate the domain shift problem in deep learning models. From the IXI dataset, which contains T1-weighted MRI scans of healthy subjects from three London hospitals, around 600 (male: 255, female: 321) images were selected for our analysis. The OASIS3 dataset, a longitudinal study focused on cognitive aging, provided a random subset of 900 MRI sessions (male: 362, female: 538) from both cognitively normal older adults and individuals with Alzheimer's disease. This dataset was specifically chosen to introduce a domain shift relative to the IXI data. All images were preprocessed to a uniform size of 128x128x128 voxels and subjected to z-score intensity normalization to ensure consistency. For training and validation, 85% of the data was used, while the remaining 15% was reserved for testing.

### B. Data Harmonization with ComBat

In order to counteract the significant domain shift that was identified during our cross-dataset validation experiments, we implemented a harmonization technique known as ComBat on all images. This methodological approach was first conceived and developed for the analysis of genomic data, where it is highly effective at correcting for variations between different laboratories and processing batches. Its utility has since been successfully demonstrated in the field of neuroimaging, where it is used to normalize data originating from disparate scanners and acquisition protocols across multiple sites. The harmonization process was applied to all T1-weighted MRI volumes from both the IXI and OASIS3 datasets, beginning with the reshaping of each 3D volume (128x128x128 voxels) into a 2D matrix by vectorizing each image into a single row. Each image in this combined dataset was then assigned a crucial batch label (0 for IXI and 1 for OASIS3) to allow the ComBat algorithm to identify the source of each data point. Utilizing the neurocombat sklearn library, the ComBat algorithm was applied to this 2D data to adjust the voxel intensities, effectively removing the inter-site variance while deliberately preserving the subject-specific variance related to biological variables like sex. Following harmonization, the adjusted 2D data were reshaped back into their original 3D volume format, and the resulting harmonized MRI volumes were saved to new directories with the remaining spatial structure, ensuring the process was applied consistently to both the training and testing splits of each dataset to prepare them for subsequent model training and evaluation.

### C. Networks Implementation and Experimental Design

We implemented and trained three modern deep learning networks specifically designed for three-dimensional data. The first was a 3D ResNet18 architecture augmented with 3D Channel Attention modules. This enhancement was integrated after each residual block, allowing the model to learn to weight the importance of its feature channels dynamically, thereby focusing its processing capacity on the most informative features within the volumetric data. The second model was a standard 3D DenseNet, chosen for its hallmark characteristic of dense connectivity, which encourages extensive feature reuse throughout the network by connecting each layer to every other layer in a feed-forward manner. The third architecture was a 3D adaptation of EfficientNet, a model renowned for its compound scaling method that systematically and uniformly balances the network's depth, width, and input resolution to achieve remarkable parameter efficiency and performance.

To thoroughly evaluate these models, four distinct training and testing strategies were employed for each one. The first strategy involved training and testing on the IXI dataset, establishing a within-domain performance baseline where the model was evaluated on data drawn from the same distribution it was trained on. Conversely, the second strategy trained the model on IXI but tested it on the OASIS3 dataset, a critical experiment designed to measure the model's cross-domain performance and its vulnerability to the domain shift between these two collections. This design was mirrored by a third strategy, which established a within-domain baseline on OASIS3, and a fourth, which tested the OASIS3-trained model on IXI. This symmetrical experimental framework was essential, as it allowed for a robust measurement of the inherent domain shift present between the two datasets (as evidenced by the performance drop in strategies two and four) and provided a means to later assess the effectiveness of any harmonization technique in mitigating this shift.

Following this initial comparative phase, the 3D ResNet18 with Attention model was selected for the final and most detailed stage of the investigation. This decision was driven by its superior performance in the within-domain baseline experiments, demonstrating a strong foundational ability to learn from the available data. Its promising baseline results made it the most suitable candidate for the subsequent harmonization experiment, where the goal was to improve its cross-domain generalization. Consequently, this model became the focus of the final harmonization procedure and the subject of the detailed analysis that followed, as it offered the greatest potential for observing and understanding the impact of addressing domain shift.

### D. Training Details and t-SNE Visualization

The models were trained using a combined Focal and Label Smoothing loss function to address specific technical challenges. Focal Loss mitigates class imbalance by reducing the contribution of easy-to-classify examples, forcing the model to focus its learning on more difficult cases. Label Smoothing improves model calibration by preventing overconfidence, replacing hard binary labels with softened values that regularize the output probabilities. This combination enhances generalization performance on unseen data. For optimization, we employed the AdamW optimizer, which improves upon standard Adam by properly decoupling weight decay from gradient updates. The learning rate was dynamically adjusted using a cosine annealing scheduler, which smoothly reduces the learning rate in a cosine pattern from its initial value to near zero. This approach allows for substantial early updates followed by finer parameter adjustments later in training. Early stopping was implemented based on validation loss to prevent overfitting and select the optimal model checkpoint.

To qualitatively assess domain shift, we extracted high-dimensional features from the penultimate layer of a model trained on OASIS3. These features capture the model's internal representation of the input data before the final classification decision. We applied t-SNE dimensionality reduction to visualize these complex features in two dimensions, preserving local cluster structures while sacrificing exact distances. We generated comparative visualizations using test sets from both IXI and OASIS3, before and after harmonization. Points were dual-coded by both dataset source and biological sex. Coloring by dataset revealed the extent of domain separation between IXI and OASIS3 samples, while coloring by sex allowed us to verify that biologically meaningful patterns were maintained throughout the harmonization process. This approach provided intuitive visual evidence of both the existing domain shift and the effectiveness of our harmonization method in aligning the datasets while preserving relevant biological information.

### III. RESULTS AND DISCUSSION

#### A. Performance Before Harmonization

Performance was evaluated using accuracy, balanced accuracy, F1-score, and Area Under the ROC Curve (AUC). Table 1 summarizes the key results before harmonization. The within-domain performance was excellent across all models (for example, ResNet18 3D with Attention achieved 95.4% accuracy on IXI and 82.8% on OASIS3). However, the cross-domain performance was severely degraded. When trained on IXI and tested on OASIS3, all models performed at chance level (~50% accuracy, 0 sensitivity), indicating a complete failure to generalize. The same was true for models trained on OASIS3 and tested on IXI, with performance dropping drastically (For instance, ResNet18 3D with Attention dropped to 54.0% accuracy). The observed F1-score of 0.00 in the cross-domain scenarios (Train on OASIS3/Test on IXI and Train on IXI/Test on OASIS3) indicates a complete model failure. This result is characteristic of a severe domain shift, where the model fails to recognize any meaningful patterns from the source domain in the target domain. Inspection of the confusion matrices revealed that the models defaulted to predicting only the majority class for all samples. This behavior confirms that the models relied entirely on dataset-specific artifacts and biases rather than learning generalizable, anatomical features relevant to sex classification. The performance effectively dropped to chance level, underscoring the profound confounding effect of inter-dataset variability.

TABLE 1: Summary of Key Results Before Harmonization

| Model | Train dataset | Test dataset | ACC | Balanced ACC | F1 Score | AUC |
|---|---|---|---|---|---|---|
| **DenseNet** | OASIS3 | OASIS3 | 0.515 | 0.510 | 0.076 | 0.734 |
| | OASIS3 | IXI | 0.172 | 0.185 | 0.25 | 0.185 |
| | IXI | IXI | 0.943 | 0.943 | 0.936 | 0.981 |
| | IXI | OASIS3 | 0.5 | 0.5 | 0 | 0.279 |
| **ResNet18** | OASIS3 | OASIS3 | 0.828 | 0.828 | 0.824 | 0.804 |
| | OASIS3 | IXI | 0.540 | 0.583 | 0.661 | 0.583 |
| | IXI | IXI | **0.954** | **0.956** | **0.95** | 0.971 |
| | IXI | OASIS3 | 0.5 | 0.5 | 0 | 0.607 |
| **EfficientNet** | OASIS3 | OASIS3 | 0.757 | 0.758 | 0.777 | 0.816 |
| | OASIS3 | IXI | 0.459 | 0.510 | 0.624 | 0.510 |
| | IXI | IXI | 0.931 | 0.938 | 0.928 | **0.997** |
| | IXI | OASIS3 | 0.5 | 0.5 | 0 | 0.721 |

## B. Performance After Harmonization

After applying ComBat harmonization, the cross-domain performance of the ResNet18 3D with Attention model improved significantly. As shown in Table 2, the balanced accuracy for "Train on IXI / Test on OASIS3" increased from ~0.50 to 0.61. Sensitivity, which was previously 0, rose to 0.64, proving the model could now correctly identify positive cases (males) from the unseen domain. Similarly, the "Train on OASIS3 / Test on IXI" scenario showed improvement. Furthermore, within-domain performance was maintained post-harmonization, confirming that the procedure preserved discriminative biological features.

TABLE 2: ResNet18 3D Performance After ComBat Harmonization

| Model | Train dataset | Test dataset | ACC | Balanced ACC | F1 Score | AUC |
|---|---|---|---|---|---|---|
| ResNet18 | OASIS3 | IXI | 0.586 | 0.625 | 0.684 | 0.855 |
| | IXI | OASIS3 | 0.61 | 0.61 | 0.621 | 0.682 |

## C. t-SNE Visualization

The t-SNE plots provide a clear visual narrative of the harmonization effect. Fig. 1 shows the features based on the trained 3D ResNet18 before harmonization, where points are clustered primarily by their dataset source (blue vs. red), with significant overlap between the clusters for males and females (green vs. purple). This indicates the model's features are dominated by domain-specific artifacts.

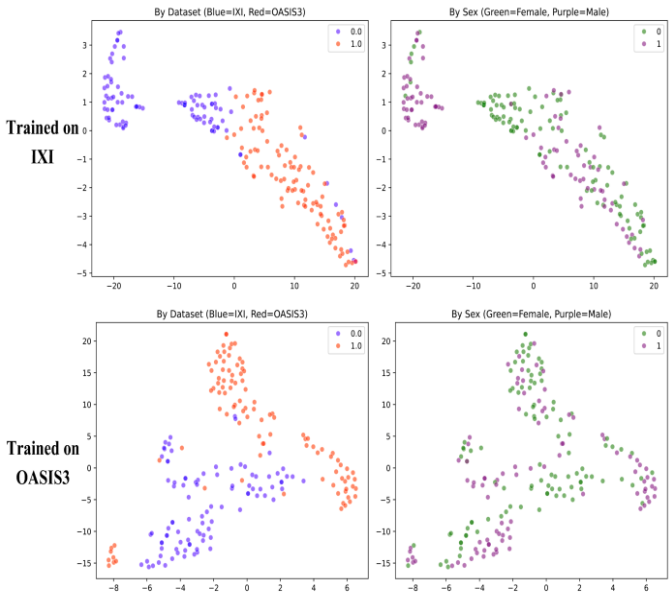

Fig. 1: t-SNE plots of deep features extracted from the test sets of IXI (Blue) and OASIS3 (Red) before harmonization, colored by Dataset (Left) and Sex (Right; Green=Female, Purple=Male).

After harmonization (Fig. 2), the dataset-specific clustering was notably reduced, with IXI and OASIS3 features showing greater overlap. At the same time, the separation based on biological sex became clearer, indicating that harmonization preserved meaningful biological variance. While the t-SNE plots provide qualitative support for this observation, future work should complement them with quantitative measures of domain alignment (for example, overlap percentage or Maximum Mean Discrepancy) to provide more objective evidence of improved generalization.

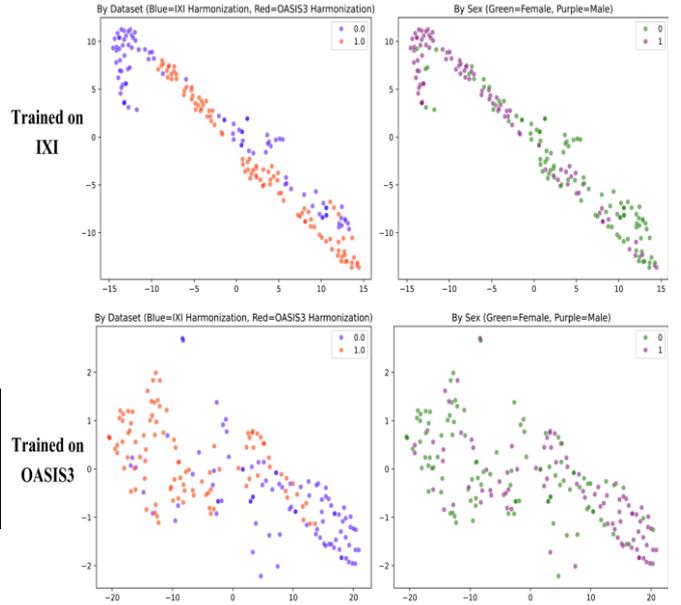

Fig. 2: t-SNE plots of deep features extracted from the test sets of IXI (Blue) and OASIS3 (Red) after ComBat harmonization, colored by Dataset (Left) and Sex (Right; Green=Female, Purple=Male).

## IV. CONCLUSION

This study provides a stark demonstration of the domain shift problem in neuroimaging AI. The catastrophic drop in cross-domain performance (Table 1) confirms that models can learn to exploit non-biological, scanner-specific noise to achieve high within-domain accuracy, a critical failure mode for clinical application. These findings emphasize that strong performance on a single dataset does not necessarily translate into robust generalization.

Our results demonstrate that ComBat harmonization is an effective strategy for mitigating this issue. The significant improvement in cross-domain balanced accuracy and sensitivity (Table 2) moves the model from a state of total failure to one of modest but meaningful performance. This represents an initial but important step toward models that rely on genuine anatomical features rather than dataset-specific artifacts. The t-SNE visualization in Fig. 1 and Fig. 2 offers a powerful explanatory tool, visually confirming that the learned feature spaces become more aligned after harmonization, which directly facilitates better generalization. These visualizations transform an abstract statistical problem into an intuitive, conceptual understanding of how harmonization works at a feature level.

However, the cross-domain accuracy (61%) remains below clinical utility. This highlights the need for more advanced domain adaptation techniques, such as adversarial training or deep domain confusion, to further improve generalization. Furthermore, ComBat requires a dataset-wise "batch" label, which may not capture more nuanced within-dataset variations. Future research should therefore focus on developing more granular and powerful harmonization techniques that can operate without coarse batch labels and bridge the remaining performance gap.

We conclusively showed that domain shift severely impacts MRI-based sex classification models. We successfully implemented ComBat harmonization, resulting in a measurable and visually verifiable reduction in domain discrepancy and a significant boost in cross-dataset generalization performance. This work provides both a methodological pathway and a compelling visual argument for prioritizing domain invariance. It underscores that data harmonization is an essential component for developing reliable and trustworthy deep learning models in neuroimaging. Ensuring generalization across scanners and sites is therefore both a technical and practical requirement for clinical translation.